\documentclass[aps,preprint,showpacs]{revtex4}

\usepackage[dvips]{graphicx}

\newcommand{\be}{\begin{equation}}
\newcommand{\ee}{\end{equation}}
\newcommand{\ba}{\begin{eqnarray}}
\newcommand{\ea}{\end{eqnarray}}
\newcommand{\beq}{\begin{equation}}
\newcommand{\eeq}{  \end{equation}}
\newcommand{\bea}{\begin{eqnarray}}
\newcommand{\eea}{  \end{eqnarray}}
\newcommand{\bit}{\begin{itemize}}
\newcommand{\eit}{  \end{itemize}}

\begin{document}

\title{ Quantum Baker Maps for Spiraling Chaotic Motion }

\author{ Pedro R. del Santoro, 
         Ra\'ul O. Vallejos \\
         and Alfredo M. Ozorio de Almeida,  }	

\affiliation{Centro Brasileiro de Pesquisas F\'{\i}sicas (CBPF), 
             Rua Dr.~Xavier Sigaud 150, 
             22290-180 Rio de Janeiro, 
             Brazil} 

\date{\today}

\begin{abstract}
We define a coupling of two baker maps through a $\pi/2$ rotation
both in position and in momentum. The classical trajectories thus
exhibit spiraling, or loxodromic motion, which is only possible
for conservative maps of at least two degrees of freedom.
This loxodromic baker map is still hyperbolic, that is, fully chaotic.
Quantization of this map follows on similar lines to other generalized 
baker maps.
It is found that the eigenvalue spectrum for quantum loxodromic baker map 
is far removed from those of the canonical random matrix ensembles.
An investigation of the symmetries of the loxodromic baker map reveals
the cause of this deviation from the Bohigas-Giannoni-Schmit conjecture.   
\end{abstract}



\maketitle

\section{Introduction}

The essential feature of classically chaotic motion is hyperbolicity,
that is both stretching and squeezing, within a confined region of phase space.
The simplest textbook examples, as far as (conservative) Hamiltonian systems 
are concerned,
are two-dimensional ($2D$) maps, which may be pictured as 
Poincar\'e sections of
motion in a $4D$ phase space. However, hyperbolicity is in no way
confined to lower dimensions. In particular, a conservative return map in four 
dimensions
may exhibit spiralling, or loxodromic motion onto a fixed point. 
This would not be allowed
in two dimensions since it would violate area conservation, but is 
compensated
in four dimensions by outward spiralling in a complementary plane.
The linearization of hyperbolic conservative maps of arbitrarity high dimension
can be decomposed into products of simple hyperbolic motion in eigenplanes
together with loxodromic motion in $4D$-subspaces \cite{Arnold}. 

Hyperbolic motion is generally desribed by horseshoe maps 
in the qualitative theory of dynamical systems \cite {GuckHol, livro}.
The idea is to describe by symbolic dynamics the successive visits 
of a trajectory to a well chosen finite set of domains of phase space. 
The problem, as far as conservative systems are concerned,
is that most orbits are not restricted to this finite set of domains, so that
the horseshoe map itself is not conservative. This is the motivation for
the develoment of a specially convenient model known as the baker map. 

The baker map displays the essential features of classically chaotic 
motion in such a simplified
form as to be almost a caricature \cite{ArnoldAvez, LichLieb}. 
It may be described as a space-filling horseshoe map
that linearizes the hyperbolic motion. Thus, essentialy chaotic 
motion is exhibited, 
which can be followed through very simple computations.
The binary symbolic dynamics propagates vertical strips in the 
unit square onto horizontal strips. 
The vertical and horizontal rectangles become narrower, 
for longer binary codes pertaining to multiple iterations of 
the baker map. 
The primary digit specifies either of the two half-squares,
so that it is responsible for a coarse-grained description of 
the motion.
The way in which horizontal strips are piled up by the mapping 
is also determined by the primary binary digit (see Fig.~\ref{fig1}). 

Higher dimensional generalizations of the baker map have recently 
been presented \cite{VSOA}.
The products of horizontal rectangles for each degree of freedom 
define
hyper-parallelepipeds that are stretched and squeezed into 
horizontal parallelepipeds.
Even though the increased dimension admits a much richer range 
of possibilities
for restacking these domains, none of these leads to spiralling 
motion.
Here we show how a rotation can be added which leads to spirals,
without violating any of the essential attributes of generalized 
baker maps.

The various quantization schemes that have been proposed for the 
baker map 
generally respect the main division of the square into a pair of 
vertical rectangles, 
which split the position states into Hilbert subspaces. 
These are mapped respectively onto a pair of momentum state 
subspaces according to the primary binary digit. 
The discrete and finite nature of the Hilbert space prevents 
the association of longer codes 
to ever thinner strips \cite{SAV}. Thus, to some extent, 
the primary digit is even more important in quantum mechanics.
 
Initially, it may have seemed that the introduction of binary 
symbols 
to describe evolving quantum systems could only serve as an 
artificial prop 
for the study of the semiclassical limit. However, the growing 
interest in quantum computation 
has brought the binary structure to the fore. 
In this context, 
the quantum baker map has reemerged as an ideal simplified model. 
Schack and Caves \cite{SchackCaves} have indeed developed
an entire class of quantum baker maps, based on the 
$2^N$-dimensional Hilbert space of $N$ qubits, 
which include the original quantizations of 
Balazs and Voros \cite{BalazsVoros} 
and Saraceno \cite{Saraceno} as special cases. 
The Schack-Caves family was recently further enlarged by Ermann and
Saraceno \cite{Ermann}, who also explained the general 
structure of all these maps.
We here adopt the Balazs-Voros-Saraceno \cite{Saraceno} map throughout. 

It is important at this stage to distinguish two alternative 
forms of classical-quantum correspondence for individual
baker maps. In the original quantizations 
\cite{BalazsVoros, Saraceno}, the classical phase space is 
two-dimensional,
a torus. It is the chaotic motion in this finite region that 
gives rise to the repulsion of the quantum quasi-energy levels, 
characteristic of the spectra of random matrices, 
according to the {\it Bohigas-Giannoni-Schmit (BGS) Conjecture} \cite{Bohigas}. 
If the quantization is chosen with a value of Planck's 
constant such that the Hilbert space has exactly $2^N$ 
states, then it is possible to reinterpret the quantum system 
to be the tensor product of $N$ two-level systems. 
This would correspond to a hypercube in a $2N$-dimensional 
phase space, but it would be stretching it to ascribe a 
classical correspondence to this quintessentially quantum system. 
We shall here keep to the original quantization schemes and, hence, 
interpret each quantum baker map as corresponding to a 
classical two-dimensional phase space, so that the coupling 
of two baker maps corresponds to a four-dimensional
phase space. The four possible combinations of primary binary 
digits, $(0,0), (0,1), (1,0), (1,1)$, then correspond to
four squares in the  classical position space, $(q_1, q_2)$ 
and hence four parallelepipeds in the unit 4-cube,
$(q_1, q_2, p_1, p_2)$ in which the classical motion is defined.   

The interaction between baker maps here involves a rotation 
of the four parallelepipeds into which the primary digits 
of both maps divide the four dimensional phase space. 
Choosing this to be of $\pi/2$, merely exchanges the 
parallelepipeds, so that the doubled binary decription 
is respected. The interesting point is that the equilibrium 
and all the classical periodic orbits become "loxodromic", 
i.e. the positions spiral outwards, while the momenta spiral 
inwards, because the eigenvalues of the stability matrix 
are general complex numbers: Neither are they real, 
nor do they lie on the unit circle. Obviously, this 
example models more closely a two dimensional motion 
for a single particle than any easily conceivable 
dynamics for a pair of one-dimensional particles. 
Loxodromic behaviour does not arise for equilibria 
of simple $p^2+ V$ Hamiltonians, but it is a generic 
possibility for the Poincar\'e map around periodic 
orbits in higher dimension. So far, not much effort 
has been made towards quantizing loxodromic motion, 
except for higher dimensional cat maps \cite{RSOA}, 
another simplified model. 

The analysis of the loxodromic baker map is preceeded 
by a brief review of the ordinary two-dimensional baker 
map and its quantization. We point out the possibilities 
for different pilings that lead to the simplest couplings 
of baker maps \cite{VSOA}. Section 3 then presents the 
loxodromic baker map and its quantization. It is shown 
in section 4 that, contrary to the generalized baker 
maps in \cite{VSOA}, the quantum eigenvalue spectrum 
does not obey a generic random matrix ensemble. The 
reason for this failure of the BGS conjecture is 
shown to be that the $\pi/2$ rotation commutes with 
the component baker map. This leads to a nontrivial 
decomposition into a tensor product of two maps, so 
that the quantum quasi-energy spectrum in this case 
is highly degenerate.

\section{Review of classical and quantum baker maps}

\begin{figure}
\includegraphics[width=12cm]{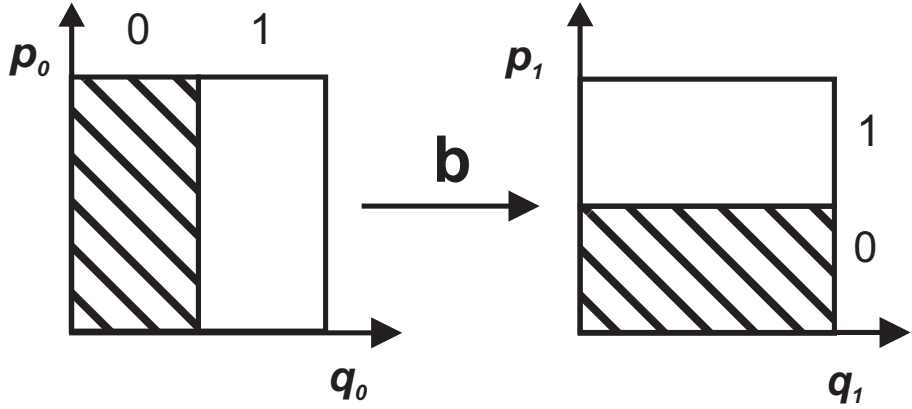}
\caption{The baker map propagates a pair of vertical rectangles 
onto a pair of horizontal rectangles. Inside each separate region
the evolution is linear.}
\label{fig1}
\end{figure}

In this section we present the well known classical and quantum
ingredients of baker maps. 
The classical baker transformation is an area preserving,
piecewise--linear map, 
${\bf b}:(p_0,q_0)\rightarrow(p_1,q_1)$ 
of the unit square 
(periodic boundary conditions are assumed) defined as
\beq
p_{1}=\frac{1}{2}(p_{0}+\epsilon_0) ~,~~~ 
q_{1}= 2q_{0}-\epsilon_0 ~;
\label{bakclas}
\eeq
where $\epsilon_{0}=[2q_{0}]$, the integer part of $2q_{0}$.  
The two regions into which the phase space is separated and 
their evolution is shown in Fig.~\ref{fig1}. 

This map is known to be uniformly hyperbolic, 
the stability exponent for orbits of period $L$ being $L\log 2$. 
Moreover it admits a useful description
in terms of a complete symbolic dynamics. A one to one correspondence
between phase space coordinates and  binary sequences,
\beq
(p,q) \leftrightarrow 
\ldots 
\epsilon_{-2} 
\epsilon_{-1} 
\cdot 
\epsilon_{0} 
\epsilon_{1} 
\epsilon_{2} 
\ldots
~~~,~ \epsilon_i=0,1 ~,
\label{symbol}
\eeq
can be constructed in such a way that the action of the map is
conjugated to a shift map, i.e., to shifting the dot to the
right in the binary sequence above.  
The symbols are assigned as follows:
$\epsilon_i$ is set to zero (one) when the $i$--th iteration of 
$(p,q)$ falls to the left (right) of the line $q=1/2$, i.e. 
$[2q_i]=\epsilon_i$. 
Reciprocally, given an itinerary
\beq
\ldots 
\epsilon_{-2} 
\epsilon_{-1} 
\cdot 
\epsilon_{0} 
\epsilon_{1} 
\epsilon_{2} 
\ldots,
\label{symbolic}
\eeq 
the related phase point is obtained through the specially
simple binary expansions
\beq
q=\sum_{i=0}^{\infty} \frac{\epsilon_i}{2^{i+1}} ~,~~~
p=\sum_{i=1}^{\infty} \frac{\epsilon_{-i}}{2^i} ~.
\label{pqsym}
\eeq
Once the dynamics has been mapped to a shift on binary sequences it is
very easy to analize the dynamical features of the map.  In particular,
periodic points are associated  to infinite repetitions of {\em finite}
sequences of symbols.

Due to its piecewise linear nature, the baker map admits a (mixed)
generating function which is a piecewise bilinear form,
\beq
W_{\epsilon_{0}}(p_{1},q_{0})=
2p_{1}q_{0} - \epsilon_{0}p_{1} -\epsilon_{0}q_{0} 
~,~~~ \epsilon_{0}=0,1 ~.
\label{genf}
\eeq
It is not defined on the whole space $p_1$--$q_0$ but on the classically
allowed domains
\beq
R_0=[0,1/2]\otimes[0,1/2] ~~~ \mbox{and} ~~~ R_1=[1/2,1]\otimes[1/2,1] ~.
\label{domains}
\eeq
Though the above generating function will be the starting point for 
quantization, 
it must be remembered that it only provides an implicit formula for
each iteration of the classical baker map. 
 
With respect to the quantum map, we will follow the original
quantization of Balazs and Voros \cite{BalazsVoros}, 
as later modified by Saraceno
\cite{Saraceno} to preserve in the quantum map all the symmetries of its
classical counterpart. In the mixed representation the baker's
propagator can be written as a $D \times D$ block--matrix ($D$ even):
\beq
\langle p_{m}|{\widehat B}_D|q_{n} \rangle= 
                \left( \begin{array}{cc}
                           G_{D/2} &      0            \\
                              0    & G_{D/2}
                       \end{array}
                \right) ~,
\label{qbdef}
\eeq
where position and momentum eigenvalues run on a discrete mesh with
step $1/D=h$ ($h$ = Planck's constant), so that
\beq
q_{n}=(n+1/2)/D ~,~~~ p_{m}=(m+1/2)/D ~,~~~ 0 \le n,m \le D-1 ~;
\label{grid}
\eeq
and $G_D$ is the antiperiodic Fourier matrix, which transforms 
from the $q$ to the $p$ basis,
\beq
G_{D}=\langle p_{m}|q_{n} \rangle=(1/\sqrt D)e^{-2\pi i D p_m q_n} ~.
\label{Fourier}
\eeq
It will be useful to consider this matrix as the $q$-representation of the Fourier
operator $\widehat{G}_D$.

The propagator for the baker map  
has the standard structure of quantized linear symplectic maps \cite{OZS},
\beq
\langle p_{m}|{\widehat B}_D|q_{n} \rangle =
\left\{ 
\begin{array}{cl}
\sqrt{2/D} \, e^{-i 2\pi D W_{0}(p_{m},q_{n})} &  
             \mbox{if} ~~ (p_{m},q_{n})\, \in \, R_0 \\
\sqrt{2/D} \, e^{-i 2\pi D W_{1}(p_{m},q_{n})} &  
             \mbox{if} ~~ (p_{m},q_{n})\, \in \, R_1 \\
0                                             &  
                                     \mbox{otherwise} ~.
\end{array}
\right.
\label{VanVleck}       
\eeq
In this quantization, only those transitions are
allowed that respect the rule $[2p_m]=[2q_n]$, a reflection of the 
classical shift property.
To be able to iterate the quantum baker map, 
the state must be brought back to the
position representation. 
This is achieved by an inverse Fourier transform,
so that the matrix (\ref{qbdef}) is multiplied by $G_D^{-1}$.  

Following Schack and Caves \cite{SchackCaves}, we can reinterpret 
this quantum map
as the evolution of $N$ qubits if the dimension of the Hilbert 
space satisfies $D=2^N$.
Then the position states can be defined as product states for 
the qubits in the basis,
\beq
|q_{n} \rangle= |\epsilon_1\rangle \otimes|\epsilon_2\rangle 
\otimes...|\epsilon_N\rangle,
\eeq
where $n$ has the binary expansion
\beq
n=\epsilon_1...\epsilon_n=\sum_{j=1}^N\epsilon_j 2^{N-j}
\eeq
and $q_n=(n+1/2)/D=0\cdot \epsilon_1...\epsilon_N 1$.
The connection with the classical baker map is specified by 
the symbolic dynamics.
The bi-infinite strings (\ref{symbolic}) that determine 
points in the unit square
are made to correspond to sets of orthogonal quantum states.
Half of the position states lie in either of the two 
rectangles, $R_0$
or $R_1$ defined in (\ref{domains}), which correspond 
respectively to $0$, 
or $1$ eigenstates of the principal qubit.

The full unitary operator for the quantum baker 
can be written explicitly as
\beq 
{\widehat B}_D={\widehat G}_D^{-1}[{\hat 1}_2 \otimes {\widehat G}_{D/2}],
\label{boperator}
\eeq 
where ${\hat 1}_2$ is the unit operator for the first bit 
and ${\widehat G}_{D/2}$
is the Fourier operator on the remaining qubits. 
The operator in the square brackets
preserves the first qubit, while evolving separately 
the remaining qubits,
within each domain $R_{\epsilon_1}$. 
It is the final
inverse Fourier operator that mixes the principal qubit 
in with the rest,
because it acts globaly on the states in both domains. 
This step is not explicitated in the mixed representation 
(\ref{qbdef}).  

\begin{figure}
\includegraphics[width=12cm]{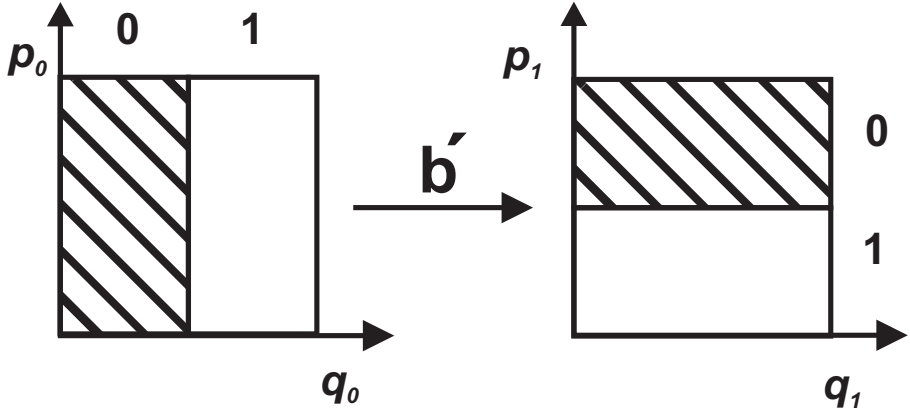}
\caption{The alternative baker map also propagates a pair of vertical rectangles 
onto a pair of horizontal rectangles, but their stacking is interchanged.}
\label{fig2}
\end{figure}
So far we have only allowed for a single possibility 
in which to stack
the rectangles in the baker transformation, but alternative 
to (\ref{bakclas}),
the classical map, ${\bf b'}:(p_0,q_0)\rightarrow(p_1,q_1)$, 
shown in Fig.~\ref{fig2},
\beq
p_{1}=\frac{1}{2}(p_{0}+1-\epsilon_0) ~,~~~ 
q_{1}= 2q_{0}-\epsilon_0 ~,
\label{bakclas'}
\eeq
has very similar properties. This variation is tantamount to
reversing the primary classical bit and it corresponds to the 
quantum map
$\widehat {B}'$, represented by the matrix:
\beq
\langle p_{m}|{\widehat B'}_D|q_{n} \rangle= 
                \left( \begin{array}{cc}
                           0 & G_{D/2}            \\
                     G_{D/2} & 0
                       \end{array}
                \right) ~.
\label{qbdef'}
\eeq
In other words, we here substitute the operator ${\hat I}_2$, 
which acted
on the principal qubit, by ${\widehat X}_2$ represented by 
the Pauli matrix:
\beq
X_2=\left( \begin{array}{cc}
                           0 & 1            \\
                           1 & 0
                       \end{array}
                \right) ~,
\eeq
so that 
${\widehat B'}_D={\widehat G}_D^{-1}[{\widehat X}_2 \otimes {\widehat G}_{D/2}]$.

Of course, we are free to substitute any other unitary operator acting on
the primary qubit, but it is only $\widehat {B}'$ that can be interpreted
classically as an equivalent alternative piling of the baker map.
We can still split up the evolution into domains equivalent to (\ref{domains}),
though with a different matching of $q_0$ and $p_1$ segments. 
The classical evolution within each domain is again determined by classical
generating functions like (\ref{genf}) which becomes the exponent of the propagator.

If we allow a finite probability for a stacking fault in the classical
baker map, the evolution acquires a random component. However,
if the piling order depends on the coarse-grained position of another
baker map, the overall motion will again be purely deterministic.
Being that each baker map is thoroughly chaotic, it will be hard to
distinguish the random motion of one of the components taken on its own
from a truly stochastic system.
In \cite{VSOA} we allowed the baker to choose between these
alternatives, depending on its interaction with another baker.
Though this led to surprisingly rich structures, no spiraling motion could
be generated in the corresponding classical map. This possibility
is now realized by rotating the domains into which the baker's
phase space is partitioned.

\section{Definition of the loxodromic baker map}

\begin{figure}
\includegraphics[width=12cm]{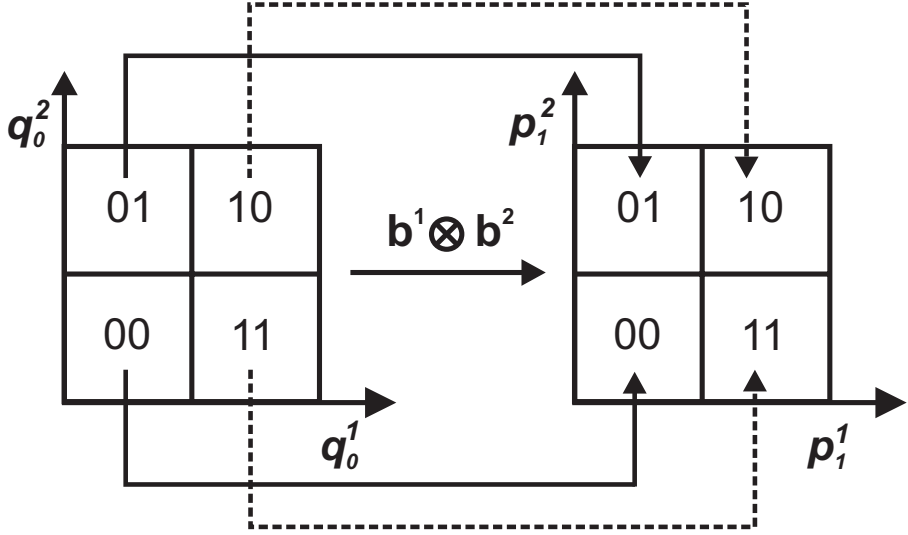}
\caption{The partition of phase space into disjoint domains 
is respected by the product of baker maps.
4D-parallelepipeds, projecting onto squares 
in the position space are transformed  onto
{\it horizontal} parallelepipeds that project onto 
squares in momentum space.}
\label{fig3}
\end{figure}

The direct product of a pair of classical baker maps, 
${\bf b}^1\otimes{\bf b}^2$, is itself
a generalized baker map defined within a 4D-hypercube, 
$(p^1, q^1)\otimes (p^2, q^2)$.
Indeed the motion is hyperbolic and the partition
of phase space into disjoint domains is again respected.
These are specified by a doubled binary code, $\epsilon_0={\epsilon_0}^1\>{\otimes}\>\>{\epsilon_0}^2$,
defining {\it vertical} 4D-parallelepipeds that 
project onto squares 
in the position space $(q^1, q^2)$, while covering 
the full available momentum space. 
The effect of a single iteration of this product map 
is to transform these domains onto
{\it horizontal} parallelepipeds that project onto 
squares in momentum space, 
as shown in Fig.~\ref{fig3}.

\begin{figure}
\includegraphics[width=17cm]{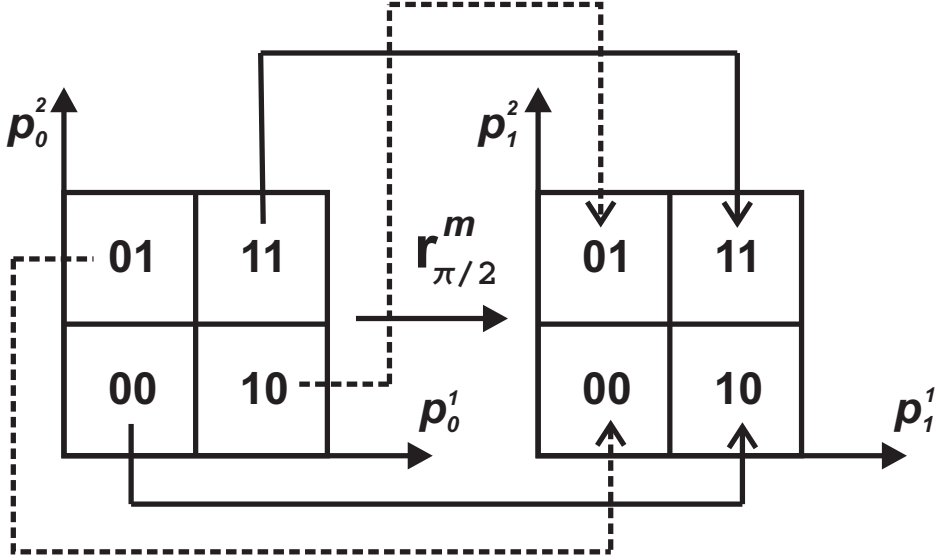}
\caption{Possible repiling of the partition of the
domains of the product of baker maps.}
\label{fig4}
\end{figure}

The main difference between this product map and the 
generalized bakers
defined in \cite{VSOA} is that in the latter case 
the symbolic mapping is no longer the identity.
One could certainly define a generalized baker by repiling
the product maps within each domain so as to obtain the 
symbolic evolution
displayed in Fig.~\ref{fig4}, though this possibility 
was not singled out in \cite{VSOA}.

\begin{figure}
\includegraphics[width=12cm]{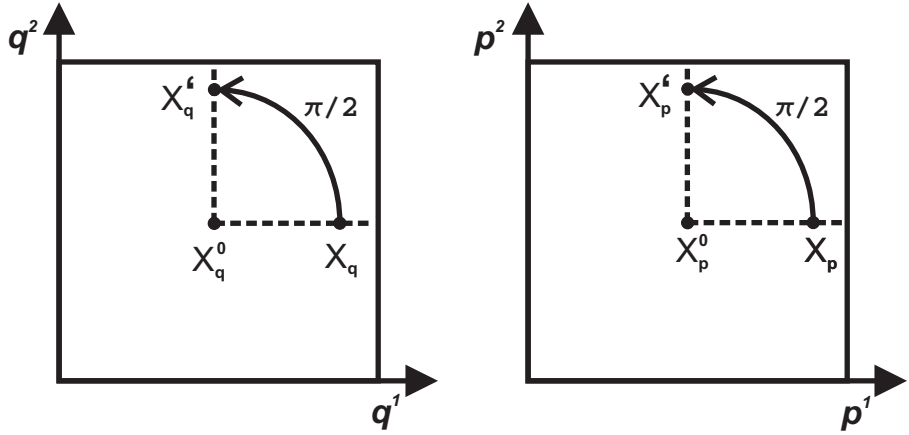}
\caption{A canonical point 
transformation, ${\bf r}_{\pi/2}^{\bf m}$, that simply rotates
both the position and the momentum spaces around the 
midpoint of the hypercube, ${\bf m}=(1/2,1/2,1/2,1/2)$,
has the same effect on the partition of the product of baker maps
as the repiling in Fig.~\ref{fig4}.}
\label{fig5}
\end{figure}

However, the same effect (as far as the symbolic evolution 
is concerned) can be obtained
by combining the product baker with a canonical point 
transformation, ${\bf r}_{\pi/2}^{\bf m}$, that simply rotates
both the position and the momentum spaces around the 
midpoint of the hypercube, ${\bf m}=(1/2,1/2,1/2,1/2)$, 
as shown in Fig.~\ref{fig5}. 

This {\em loxodromic baker map}, 
${\bf r}_{\pi/2}^{\bf m}\circ [{\bf b}^1 \otimes {\bf b}^2]=
 {\bf b_L}:(p_0,q_0) \rightarrow (p_1,q_1)$ 
is given explicitely by
\bea
{p_1}^1 & = & 1-{1\over 2}{p_0}^2-{1\over 2}{\epsilon_0}^2 \\
{p_1}^2 & = & {1\over 2}{p_0}^1-{1\over 2}{\epsilon_0}^1   \\ 
{q_1}^1 & = & 1- 2{q_0}^2-{1\over 2}{\epsilon_0}^2         \\
{q_1}^2 & = & 2{q_0}^1-{1\over 2}{\epsilon_0}^1.
\eea

\begin{figure}
\includegraphics[width=12cm]{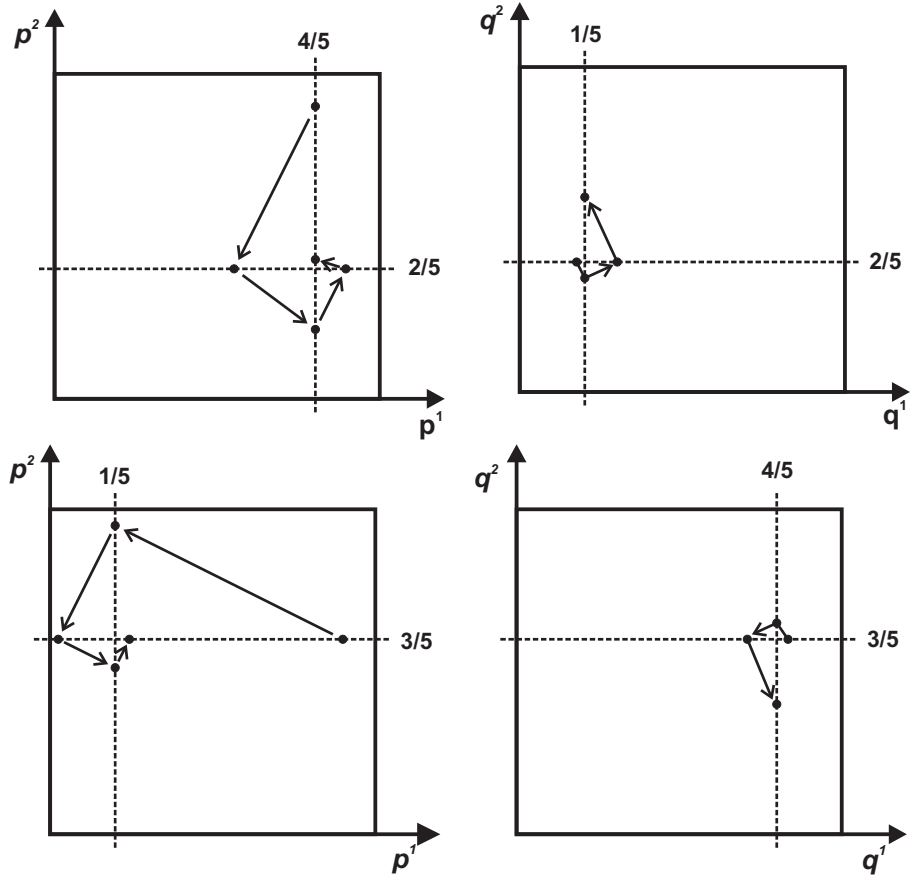}
\caption{Typical classical trajectories of the loxodromic baker map 
spiral out of the fixed points in the q-plane and into the fixed points
in the p-plane. Every four iterations they return to the same radial line,
because the rotation angle is $\pi/2$.}
\label{fig6}
\end{figure}

This is no longer a product map within each separate domain and 
Fig.~\ref{fig6} shows how an orbit spirals onto a fixed point in 
momentum space
and spirals outwards in position space.

The loxodromic baker map can also be described by symbolic dynamics.
This now has a four letter alphabet, two binary digits, 
instead of one for the ordinary baker map. Furthermore,
the relation between the symbolic description and the phase space trajectories
is not so direct. The details may be obtained in \cite{Santoro}.

The quantum version of the loxodromic baker map is defined in
strict correspondence with the foregoing classical construction. 
This is always possible for linear classical evolutions, i.e.
we can define a quantum unitary transformation, $\widehat{R}_{\pi/2}^{\bf m}$,
which corresponds to the classical transformation, ${\bf r}_{\pi/2}^{\bf m}$,
so that, in combination with the definition of the usual baker map 
(\ref{boperator}), we then stipulate that
\beq
{\widehat B}_L = \widehat{R}_{\pi/2}^{\bf m}\>\> {\widehat B}_D \otimes {\widehat B}_D.
\label{qlox}
\eeq

The only thing that is now missing is the full definition 
of the rotation operator in the above formula. 
Recalling that the classical rotation is about
the midpoint ${\bf m}$ in phase space, it follows that,
if the rotation operator about the origin by an angle $\alpha$ is
\beq
\widehat{R}_{\alpha}^0= 
\exp \left[-{i \alpha}
({\widehat p}_2 {\widehat q}_1 - 
{\widehat q}_2 {\widehat p}_1) \right],
\eeq
then
\beq
\widehat{R}_{\alpha}^{\bf m}= 
{\widehat T}_{\bf m} \> \widehat{R}_{\alpha}^0 \>{\widehat T}_{-\bf m},
\eeq
where ${\widehat T}_{\xi}$ is the quantum operator corresponding to a uniform
translation (or displacement) by the phase space vector $\xi$ (see e.g.\cite{ozorio98}).

A considerable simplification is possible for the $\pi/2$ rotation, 
which is the only one allowed for a generalized baker map. 
In this case, it is easy to verify
that the classical rotation about the midpoint, $\bf m$, can be written
as a composition
\beq
{\bf r}_{\pi/2}^{\bf m} = {\bf t}_{\eta} \circ {\bf r}_{\pi/2}^0,
\eeq
where ${\bf t}_{\eta}$ is the classical translation by the vector
$\eta=(1,0,1,0)$ and ${\bf r}_{\pi/2}^0$ the $\pi/2$ rotation around the 
origin. 
Therefore, we can quantize this form directly
so as to obtain
\beq
\widehat{R}_{\pi /2}^{\bf m}= {\widehat T}_{\eta}\> \widehat{R}_{\pi/2}^0.
\eeq

\section{Symmetries and level repulsion}

The  celebrated {\it BGS conjecture} predicts the 
correspondence between chaotic classical motion 
and quantum level repulsion. Specifically, for quantum maps, 
the eigenvalue spectrum
of the unitary operator should have fluctuations in agreement 
with the canonical
CUE, or COE ensembles of random matrix theory (RMT). 
Exceptions do arise because of symmetries,
or even more subtle arithmetic anomalies which may be hidden 
in the definition of the map.
However, the BGS conjecture has become so well 
established that its negation
is taken as a sure sign that hidden structure should be 
sought out.

In relation to the baker map, it should be noted that the
original quantization by Balazs and Voros \cite{BalazsVoros} 
did not conform properly to the conjecture. 
Saraceno then discovered that this form of quantization did not 
incorporate an exact classical symmetry and proposed a 
modification \cite{Saraceno}. 
The individual symmetry classes of this new map are satisfactory 
examples of the BGS conjecture as can be verified in \cite{OCT}.

\begin{figure}
\includegraphics[width=12cm]{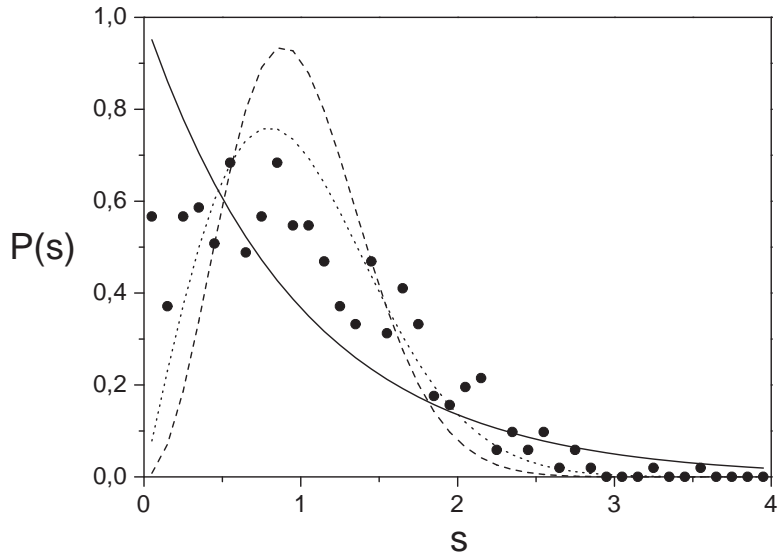}
\caption{Distribution of level spacings --normalized to unit average-- 
for the repiled baker maps (full dimension equal to 512; big dots).
We also display the predictions of random matrix
theory (dotted line: COE, dashed: CUE) and the Poisson distribution (full line).}
\label{fig7}
\end{figure}

\begin{figure}
\includegraphics[width=12cm]{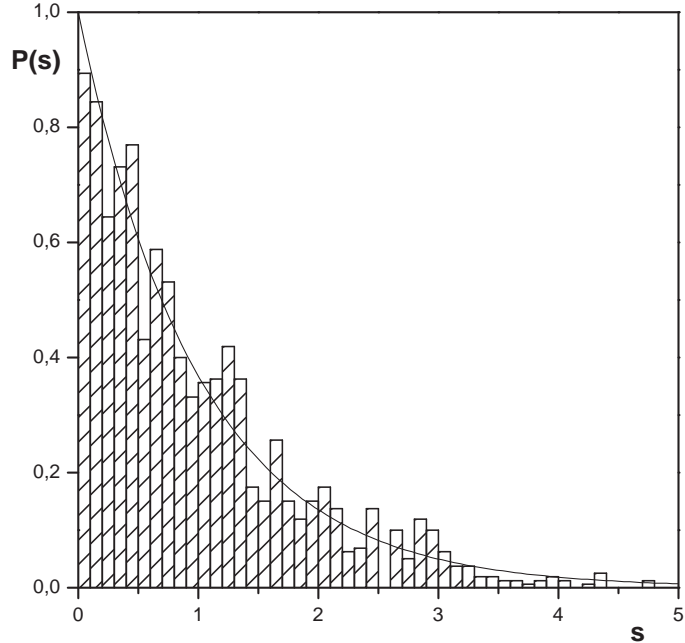}
\caption{Histogram of level spacings for eigenangles of a loxodromic 
baker map with dimension equal to 1296.
The smooth curve corresponds to the Poisson spectrum.}
\label{fig8}
\end{figure}

Even though the direct product of two baker maps is certainly
a chaotic classical system, its quantization does not
satisfy the BGS conjecture. Indeed, the eigenvalues will
be the products of the baker eigenvalues, leading to degeneracies
instead of level repulsion. In general, the addition of some 
interaction between the two maps, 
while taking the precaution to preserve the hyperbolicity of the 
classical motion, should lead again to an RMT spectrum. 
In the case of the generalized ``repiled" baker maps studied in 
\cite{VSOA} the level spacing distribution, shown in Fig.~\ref{fig7},
does not comply with the BGS conjecture.
Its intermediate nature between a Poisson spectrum and a COE spectrum
suggests the nearness of a symmetry, as was found for the original 
Balasz-Voros quantization of the baker map.
However, Fig.~\ref{fig8} clearly rules this possibility out in the case
of the loxodromic baker map.
The Poisson profile indicates independent samples and, furthermore, 
there are many degeneracies, just as with the simple product map. 
Therefore we must seek out the symmetry that is responsible.
It turns out that this involves the reflection symmetry of the 
classical baker map,
${\bf R}: (q,p)\rightarrow R(q,p)=(1-q, 1-p)$, included in the 
Saraceno quantization \cite{Saraceno}. 
Of course, both factors of the product map, 
${\bf b}^1\otimes{\bf b}^2$,
commute with the respective reflections, 
${\bf R}_1$ and ${\bf R}_2$.
Another obvious symmetry of the product map is the permutation,
${\bf P}: (q_1, p_1, q_2, p_2) 
\rightarrow P(q_1, p_1, q_2, p_2)=(q_2, p_2, q_1, p_1)$.
Combining these, we then obtain:
\beq
 {\bf R}_1 P(q_1, p_1, q_2, p_2)= R_1(q_2, p_2, q_1, p_1)
=(1-q_2, 1-p_2, q_1, p_1)
= {\bf r}_{\pi/2} (q_1, p_1, q_2, p_2).
\eeq
Thus, we find that the special nature of the $\pi/2$ rotation, 
that was chosen because
it respects the geometrical divisions of the baker map, can 
now be reinterpreted
as a particular product of the symmetries of 
${\bf b}^1\otimes{\bf b}^2$.

It is then neither surprising, nor hard to prove, 
that ${\bf r}_{\pi/2}$ itself
is a symmetry of the classical and the quantum product baker map. 
The consequence is that $\widehat{R}_{\pi/2}$
and ${\widehat B}_D \otimes {\widehat B}_D$ have common eigenfunctions and 
they must share these
with ${\widehat B}_L$, as defined by (\ref{qlox}).
Thus, if we divide the spectrum of the product map into the symmetry 
classes of the quantum rotation 
${\widehat{ R}}_{\pi/2}$,
with eigenvalues $\{+1, -1, +i, -i\}$, the eigenvalues of 
${\widehat B}_L$ in each
class are just the corresponding multiple of the eigenvalues 
of the quantized product map.
In conclusion, the loxodromic baker map belongs to the class 
of quantized chaotic maps
which fail to satisfy the BGS conjecture because of a symmetry.

\section{Discussion} 

The quantization of spiraling chaotic motion has not received the 
attention
it deserves, because, so far, work has concentrated on 
two-dimensional maps.
For systems with many degrees of freedom, it must be borne in 
mind that
Poincar\'e maps will generally exhibit spiraling behaviour along with the 
more familiar
features studied in two-dimensional maps. 
The complexity of higher dimensional motion
encourages the search for specially simple models with which to study 
the classical-quantum correspondence
and the semiclassical limit. The generalized baker map presented here 
was an
obvious candidate, but the restriction to rotations of $\pi/2$,
amounting roughly to a permutation of the variables 
between the two coupled maps, turns out to be too severe.
The commutation of this rotation with the baker product map leads to a 
totally unsatisfactory
eigenvalue spectrum for a chaotic quantum map.
This is in sharp contrast to even simpler couplings developed
for baker maps in \cite{VSOA}

Another possibility that has also been investigated are loxodromic cat 
maps, i.e.
linear maps on the 4-torus \cite{RSOA}. 
There is then a richer range of possibilities,
but still there are restrictions to the distribution of eigenvalues 
common to all cat maps.
These arise from the necessity of quantum cat maps to be periodic, 
albeit the period may be very high.    
Presumably, the search for simple loxodromic quantum models will only 
heat up when
a loxodromic algorithm is invented for quantum information theory.


\end{document}